\documentclass{svproc}
\usepackage[utf8]{inputenc}
\usepackage{graphicx}
\usepackage{amsmath}
\usepackage{amsfonts}
\usepackage{amssymb}
\usepackage{makeidx} % For index
\makeindex % For index
\usepackage{hyperref}
\usepackage{cleveref}
\usepackage{booktabs}

\usepackage{tikz}
\usetikzlibrary{arrows.meta, positioning}

\graphicspath{{figs/}} % Specify graphics path

\title{Threat Modeling for Enhancing Security of IoT Audio Classification Devices under a Secure Protocols Framework}
\titlerunning{Threat Modeling for IoT Audio Devices}
\author{Sergio Benlloch-Lopez\and Miquel Viel-Vazquez \and Javier Narnajo-Alcazar \and
Jordi Grau-Haro \and Pedro Zuccarello}
\authorrunning{Sergio Benlloch-Lopez et al.} % abbreviated author list (for running head)
\institute{Instituto Tecnologico de Informatica (ITI), Valencia, Spain\\
\email{\{sbenlloch, mviel, jnaranjo, jgrau, pzuccarello\}@iti.es}
}

\begin{document}

\maketitle

\begin{abstract}
The rapid proliferation of IoT nodes equipped with microphones and capable of performing on-device audio classification exposes highly sensitive data while operating under tight resource constraints. To protect against this, we present a defence-in-depth architecture comprising a security protocol that treats the edge device, cellular network and cloud backend as three separate trust domains, linked by TPM-based remote attestation and mutually authenticated TLS 1.3. A STRIDE-driven threat model and attack-tree analysis guide the design. At startup, each boot stage is measured into TPM PCRs. The node can only decrypt its LUKS-sealed partitions after the cloud has verified a TPM quote and released a one-time unlock key. This ensures that rogue or tampered devices remain inert. Data in transit is protected by TLS 1.3 and hybridised with Kyber and Dilithium to provide post-quantum resilience. Meanwhile, end-to-end encryption and integrity hashes safeguard extracted audio features. Signed, rollback-protected AI models and tamper-responsive sensors harden firmware and hardware. Data at rest follows a 3-2-1 strategy comprising a solid-state drive sealed with LUKS, an offline cold archive encrypted with a hybrid post-quantum cipher and an encrypted cloud replica. Finally, we set out a plan for evaluating the physical and logical security of the proposed protocol.
\end{abstract}

\begin{keywords}
IoT security, threat modeling (STRIDE), TPM, audio classification, post-quantum cryptography, Machine Learning, Edge Computing.
\end{keywords}

%!TEX root = ../main.tex

\section{Introduction}
The proliferation of Internet of Things (IoT) devices has revolutionized various aspects of daily life, from smart homes \cite{Vardakis2024Review} to industrial automation \cite{Wang2023Security}. Among these devices, IoT audio classification units are gaining significant traction. They interpret ambient sounds for anomaly detection \cite{LoScudo2023Audio}, security monitoring \cite{Al-Khalli2023Real-Time}, and elder care \cite{Yang2025RealTimeASR}. These units often have limited computational and energy resources. They also process sensitive audio data, making them prime targets for cyberattacks. Their inherent design and deployment vulnerabilities require a robust security framework \cite{Ammar2018InternetOT}.

Awareness of IoT security is increasing. However, existing solutions often fail to cover the unique challenges faced by audio classification devices \cite{Kumar2021Survey}. Traditional protocols, crafted for more powerful hardware, are usually too
resource-intensive for these devices. Audio data is subjected to threats such as eavesdropping, data manipulation, and
impersonation. These threats require tailored measures that generic IoT paradigms do not provide. This gap exposes IoT audio classification systems to serious risks. Potential consequences include compromised privacy, degraded system integrity,
and interrupted operation.

\begin{figure}[!ht]
    \centering
    \includegraphics[width=1\linewidth]{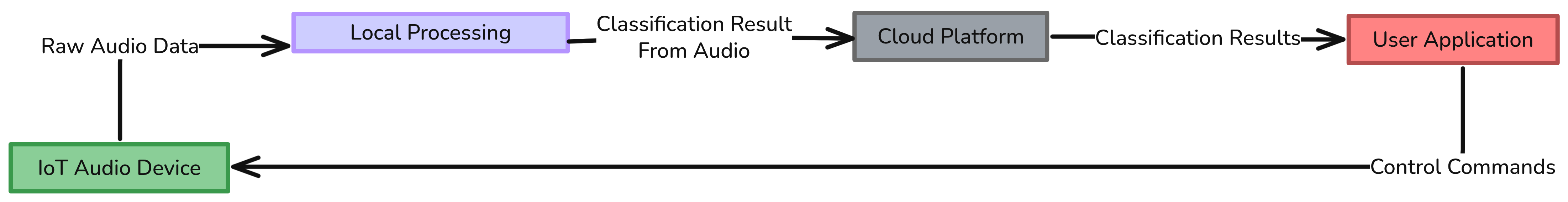}
    \caption{Overview of the IoT audio classification system architecture.}
    \label{fig:architecture}
\end{figure}

This paper addresses these critical security challenges by proposing a comprehensive threat modeling approach integrated with the design of secure communication protocols for IoT audio classification devices. The high-level architecture of our proposed system is illustrated in Fig. \ref{fig:architecture}.

The system starts with an IoT audio device (green, edge node), which captures raw audio data from the environment. This data is sent to a local processing unit (purple, on-device processing), where the audio is preprocessed and processed with a model at the edge. Instead of transmitting the raw audio to the cloud, only the processed results are sent to the Cloud Platform (gray, cloud), which reduces bandwidth usage and addresses privacy concerns. The cloud stores the classifications and end users can access the results through a user application (red, user interface), typically as a real-time dashboard. The diagram illustrates the information flow and distribution of processing tasks, highlighting the division between processing at the edge and analysis in the cloud, and the feedback loop that allows user commands to reach the device.
\\\\
To address these gaps, we offer four main contributions:
\begin{itemize}
    \item (C1) Threat-to-control traceability. A STRIDE-driven attack tree and a coverage matrix that map each identified risk to concrete controls across device, link, and backend trust boundaries.
    \item (C2) Attestation-gated data-at-rest pattern. A design pattern where the cloud releases a short-lived LUKS key only after a TPM quote satisfies policy (PCRs, geofence/time). The pattern is specified as an interface contract and sequence diagrams—no new cryptography is claimed.
    \item (C3) Minimal mTLS control-plane profile. A request/response profile over TLS 1.3 with mutual auth, certificate stapling, and tokenized policy decisions tailored to intermittent 4G links; we define message types, timeouts, and reconnect behavior as design guidance.
    \item (C4) PQC-hybrid upgrade path. A migration blueprint (e.g., Kyber for KEM, Dilithium for signatures) indicating where hybrids fit in bootstrapping, transport, and backups; algorithm choices are illustrative, not prescriptive.
    \item (C5) Evaluation plan \& acceptance criteria. A set of implementation-agnostic metrics and experiments (latency budget, CPU/energy overhead of attestation/unlock, recovery/rotation drills) to be executed by future prototypes to validate the architecture.
\end{itemize}

\textbf{Organization:} Section 2 reviews related work and positions our approach. Section 3 presents the STRIDE-based threat model and attack tree. Section 4 details the protocol (attestation, mTLS, LUKS unlocking, model signing, and PQC hybrids). Section 5 defines our evaluation methodology. Section 6 discusses trade-offs, limitations, and operational lessons. Section 7 concludes and outlines future work.

To the best of our knowledge, this is the first public work that combines the use of sensors, TPM, and advanced cryptography to protect models deployed at the edge.
%!TEX root = ../main.tex

\section{Related Work}

\begin{figure}[!ht]
    \centering
    \includegraphics[width=1\linewidth]{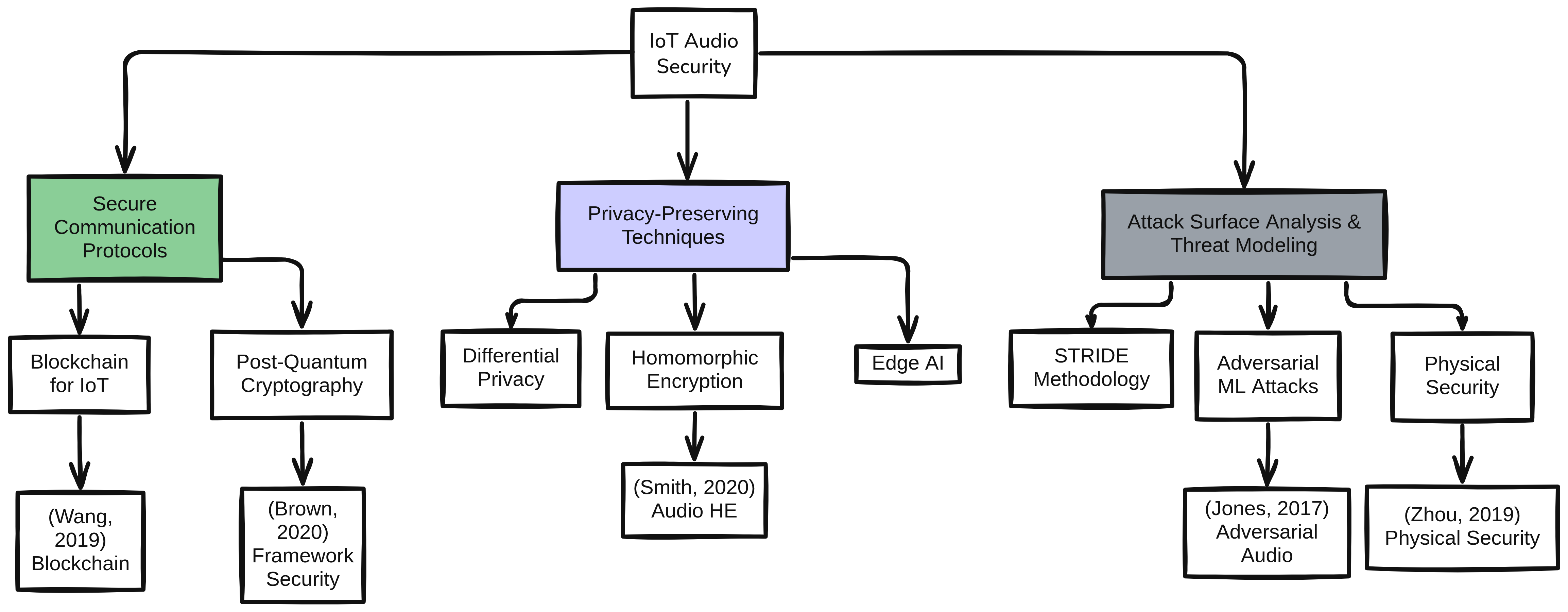}
    \caption{ Taxonomy of IoT Audio Security. The diagram categorizes IoT audio security strategies into three main areas: Secure Communication Protocols (green), Privacy-Preserving Techniques (purple), and Attack Surface Analysis \& Threat Modeling (grey). Each branch presents representative methods or references, highlighting the layered approach needed for secure IoT audio systems.}
    \label{fig:related_work_taxonomy}
\end{figure}

Security in IoT audio classification devices is a rapidly evolving field, with recent research focusing on various aspects of threat mitigation and secure protocol design \cite{Turchet2020Internet}. This section reviews state-of-the-art contributions, categorizing them by their primary focus: attack surface analysis, secure communication protocols, and privacy-preserving techniques, as summarized in Fig. \ref{fig:related_work_taxonomy}.

\subsection{Attack Surface Analysis and Threat Modeling}
Recent studies have highlighted the expanded attack surface of IoT devices, particularly those handling sensitive audio data. For instance, Pin-Yu Chen and Sijia Liu investigated in a novel adversarial attacks targeting IoT audio classification models \cite{Chen2024Adversarial}, demonstrating how manipulated audio inputs can lead to misclassification or unauthorized command execution. Similarly, W. M. A. B. Wijesundara et al. examined the vulnerabilities in the firmware of IoT audio devices, identifying potential exploits in \cite{Brown2024FirmwareSecurity}. Threat modeling frameworks, such as STRIDE, have been adapted for IoT contexts, as shown by C. Ellerhold et al. in  \cite{Jones2023ThreatModeling}, who applied a comprehensive threat model to IoT ecosystems, identifying key areas of concern for audio-enabled devices. The work in \cite{Zhao2024PhysicalSecurity} further extended this by reviewing physical security measures for IoT devices, providing insights into protecting against physical tampering.

\subsection{Secure Communication Protocols}
Ensuring secure communication is paramount for IoT audio devices, given their often-remote deployment and reliance on cloud services. Traditional TLS/DTLS (Datagram Transport Layer Security) protocols, while robust, can be resource-intensive for constrained devices. Haroon et al. in \cite{Kim2025LightweightDTLS} proposed a lightweight DTLS 1.3 implementation optimized for resource-constrained IoT audio devices, demonstrating reduced handshake overheads suitable for intermittent audio data transmission. Furthermore, Liu et al. in \cite{Liu2025PostQuantum} investigated the feasibility of postquantum cryptography in IoT audio devices, assessing its potential for future-proofing audio data security.

\subsection{Privacy-Preserving Techniques}
Beyond communication security, the privacy of audio data itself is a significant concern. Homomorphic encryption, as applied by \cite{Taylor2021Homomorphic}, allows for computation on encrypted audio data, enabling cloud-based classification without exposing raw audio. Differential privacy mechanisms have also been explored to protect user identities from audio datasets, with J. G. Bourrée in \cite{Garcia2025Privacy} proposing privacy-preserving techniques for audio data in IoT environments. AI-powered security, including anomaly detection for securing IoT audio streams, has been investigated by \cite{Miller2025AIAnomaly}, contributing to both security and privacy. The increasing adoption of Edge AI, as discussed by H. Smith et al. in \cite{Smith2024EdgeAI}, also enhances privacy by processing audio data locally on devices, reducing the need to transmit sensitive information to the cloud.

This body of work provides a foundation for our proposed approach, which integrates a comprehensive threat model with a tailored secure protocol design to address the specific challenges of IoT audio classification devices.
%!TEX root = ../main.tex

\section{Threat Model}\label{sec:threatmodel}

This section details the threat model for an IoT audio classification system comprising an edge IoT audio device with embedded Machine Learning (ML) capabilities, a 4G (Fourth Generation) cellular connection for data transmission, and a backend API server interacting with a PostgreSQL database. The IoT audio device operates in a primary-secondary communication model, initiating all communications with the API server for data transmission, inferences, monitoring probes, and model updates via a secure HTTPS API.

The API server is protected by a Web Application Firewall (WAF), following NIST SP 800-228 \cite{NIST}, and leverages HTTPS for all communications. We apply the STRIDE (Spoofing, Tampering, Repudiation, Information Disclosure, Denial of Service, Elevation of Privilege) \cite{STRIDE}, threat modeling methodology to systematically identify potential threats across these components.

\subsection{Attack Tree Analysis}
Attack trees are a structured way to visualize and analyze security risks, supporting systematic threat modeling. They help identify critical vulnerabilities, prioritize defenses, and anticipate attacker behavior by mapping all feasible attack strategies against the system.

\begin{figure}[!ht]
    \centering
    \includegraphics[width=1\linewidth]{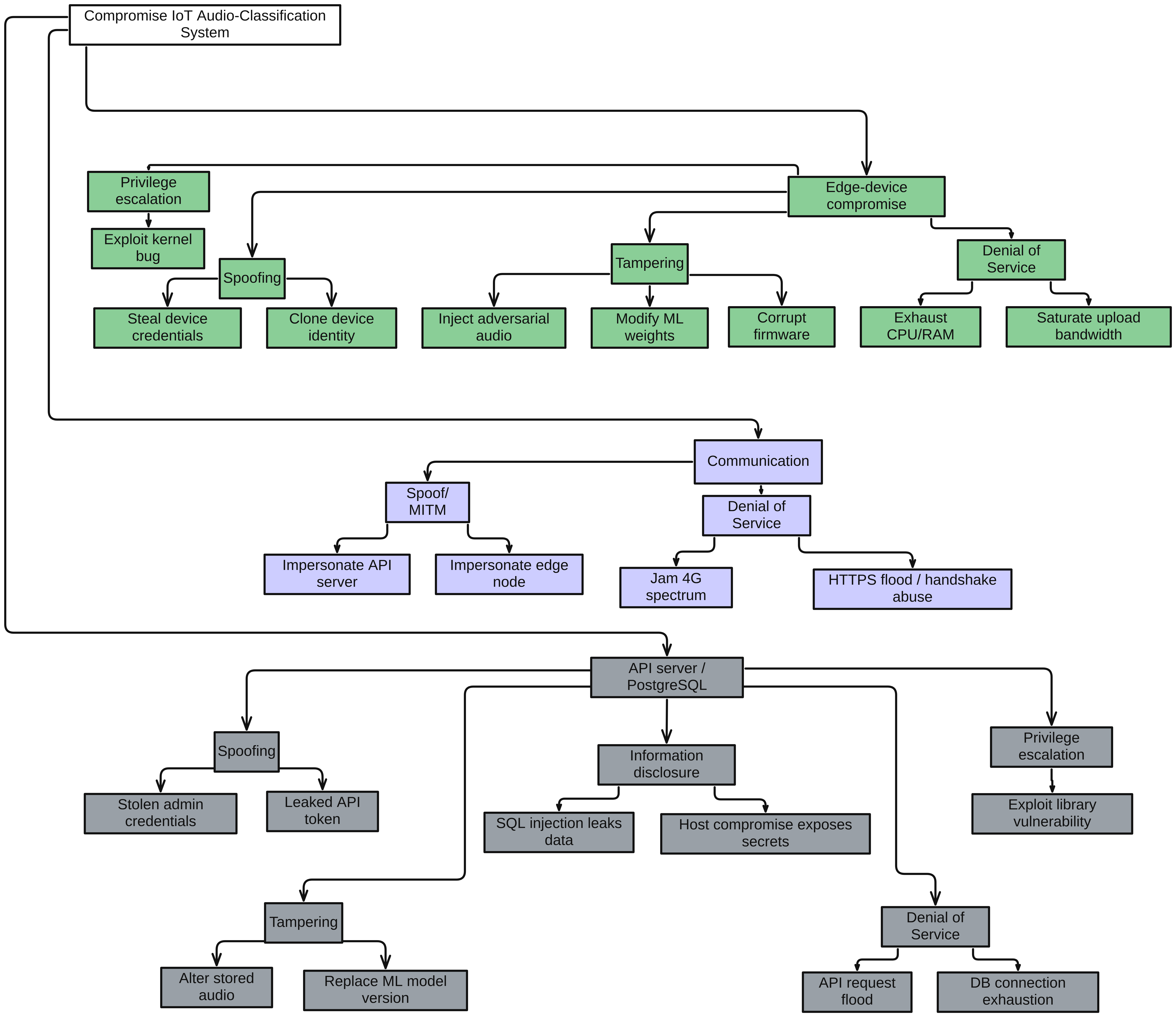}
    \caption{Attack Tree for IoT Node Compromise. This diagram illustrates possible attack paths to compromise an IoT audio classification system. Each branch represents a different attack scenario, breaking down complex threats into actionable steps. The tree begins with the main goal, system compromise, and decomposes it into sub-goals such as edge device compromise (green), communication attacks (purple), and backend/API server threats (grey). Each colored block groups related attack vectors, including privilege escalation, spoofing, tampering, denial-of-service, and more specific exploits like adversarial audio injection or API impersonation.}
    \label{fig:attack_tree}
\end{figure}

To further illustrate the potential attack paths, we present an attack tree in Fig~\ref{fig:attack_tree}, focusing on the objective of compromising the IoT audio device.

\subsection{Edge Device}

The IoT audio device is the primary data collection and initial processing point. It features enhanced security measures including ciphered partitions, processes running with Secure Boot, and protected \texttt{/audio} and \texttt{/models} partitions. We define two partitions to control the size of the saved audios and saved models, and to create an internal division within the system. This also allows us to have independent keys between the partitions. Deciphering reboots are managed using a remote API and LUKS (Linux Unified Key Setup) ciphering systems.

\begin{itemize}

    \item \textbf{Spoofing:} An attacker could impersonate a legitimate IoT audio device to inject malicious data or commands into the system. This could involve cloning device identities or compromising authentication credentials.
    
    \item \textbf{Tampering:} The ML model or the audio data collected on the device could be altered. This includes injecting adversarial audio samples, modifying model parameters to misclassify, or corrupting firmware during updates.
    
    \item \textbf{Repudiation:} A compromised IoT audio device could deny having sent specific data or executed certain commands, making accountability difficult.
    
    \item \textbf{Information Disclosure:} Sensitive audio data or ML model details could be exfiltrated from the device. This is particularly critical if the raw audio contains personally identifiable information.
    
    \item \textbf{Denial of Service:} An attacker could overload the limited resources of the IoT audio device, preventing it from collecting or process audio or transmitting data. This could be achieved through excessive data ingress or resource-intensive computations.
    
    \item \textbf{Elevation of Privilege:} An attacker gaining low-level access could escalate privileges to gain full control over the device, enabling arbitrary code execution or complete system compromise.
    
\end{itemize}

\subsection{Communication}

The 4G connection serves as the communication backbone between the IoT node and the API server. With the adoption of HTTPS, several threats are inherently mitigated:

\begin{itemize}

    \item \textbf{Spoofing:} Mutual authentication via X.509 (ITU-T X.509) certificates within HTTPS prevents an attacker from impersonating either the IoT node or the API server, effectively mitigating man-in-the-middle attacks that rely on spoofing the communication endpoints.

    \item \textbf{Tampering:} The integrity checks provided by TLS within HTTPS ensure that any modification of data in transit between the IoT node and the API server will be detected, preventing data manipulation and ensuring the authenticity of messages.

    \item \textbf{Repudiation:} While HTTPS provides some level of non-repudiation for the transport layer, an attacker could still attempt to deny having sent specific data or commands at the application layer, especially if the IoT audio device is compromised. To mitigate this, application-level digital signatures on data payloads, combined with secure logging and timestamping, are crucial. This ensures that each message's origin and content can be cryptographically verified, even if the device itself is later compromised or attempts to deny its actions.
        
    \item \textbf{Information Disclosure:} HTTPS, using TLS (Transport Layer Security), encrypts all data in transit, significantly reducing the risk of sensitive information being intercepted and decrypted by adversaries, even if the underlying 4G network has vulnerabilities. This provides strong end-to-end encryption for the application layer.
    
    \item \textbf{Denial of Service:} While jamming or flooding the cellular frequency remains a physical threat. The primary-secondary model, where the IoT node initiates connections, can help manage resource consumption on the server side by only processing requests from authenticated clients.

    \item \textbf{Elevation of Privilege:} If an adversary steals or forges valid client certificates, session tokens, or API keys during transport, they can escalate from unauthenticated actor to privileged API role, issuing administrative commands, deploying malicious firmware, or accessing restricted data. Mitigations include hardware-bound private keys on the device, short-lived credentials, strict role-based access control, and continuous monitoring for anomalous privilege use.

\end{itemize}

\subsection{Server}

The backend infrastructure is responsible for receiving, storing, and further processing data, as well as serving ML model updates and configurations. With the shift to HTTPS for all node-to-server communication, the following considerations apply:

\begin{itemize}

    \item \textbf{Spoofing:} HTTPS with mutual authentication significantly reduces the risk of an attacker impersonating the API server to IoT nodes or vice versa. However, impersonation of legitimate clients (e.g., administrators) to the API server remains a concern if their credentials are compromised.
    
    \item \textbf{Tampering:} Data integrity is largely ensured during transit by HTTPS. However, data stored in the PostgreSQL database or processed by the API could still be altered if the server itself is compromised. This includes modifying historical audio data, ML model versions, or user configurations.
    
    \item \textbf{Repudiation:} The use of HTTPS and API calls, especially when combined with proper logging and authentication, can strengthen non-repudiation. The API server can log authenticated requests and their payloads, making it harder for a compromised node or client to deny having sent specific data or executed certain commands.
    
    \item \textbf{Information Disclosure:} While HTTPS encrypts data in transit, sensitive data (e.g. raw audio, processed features, user metadata, ML model weights, API keys, database credentials) could still be leaked from the server or database if the server is compromised. SQL injection remains a common vector here, emphasizing the need for robust input validation.
    
    \item \textbf{Denial of Service:} The API server or PostgreSQL database could be overwhelmed by excessive requests, preventing legitimate operations. While HTTPS itself does not prevent DoS, the primary-secondary API model allows for rate limiting and stricter access controls at the API gateway level, which can help mitigate DDoS attacks. However, resource exhaustion from legitimate but excessive requests (e.g., too frequent polling for updates) needs to be managed.
    
    \item \textbf{Elevation of Privilege:} An attacker gaining access to the API server or database with limited permissions could exploit vulnerabilities to gain administrative control, leading to full system compromise. This threat is independent of the communication protocol and requires strong access control, secure coding practices, and regular security audits.

\end{itemize}

This comprehensive threat model informs the design of secure protocols and architectural considerations to mitigate these identified risks.
%!TEX root = ../main.tex

\section{Protocol Design}\label{sec:protocol_design}

This section outlines the secure protocol design for the IoT audio classification system, focusing on mitigating the threats identified in the previous section. The design combines established security standards with optimisations for IoT environments with limited resources and the specific requirements of IoT audio devices enabled by machine learning.

\subsection{Secure Communication between IoT Node and API Server}

Given the 4G cellular connection, a robust and efficient communication protocol is crucial. The IoT audio device acts as a client, initiating all communications with the API server (primary-secondary model) for data transmission, updates, and other necessary operations. We propose a secure channel established using HTTPS, mutual authentication, and API-based communication. One of the main reasons for using HTTPS instead of MQTT or another IoT protocol is the requirement for a request-response scheme. This is necessary in order to trigger updates or changes to online functionalities of the node via API responses. In the event of scalability issues, we propose adapting the scheme to use a CoAP broker while maintaining the API.

\subsection{Data Security and Integrity}

\begin{itemize}
    \item \textbf{End-to-End Encryption:} Sensitive audio features are encrypted on the IoT audio device before transmission and decrypted only at the API server, adding a second protection layer in case the transport channel is compromised.
    \item \textbf{Data Integrity Checks:} Cryptographic hash functions ensure the integrity of audio data and ML model updates, any tampering during transmission or storage is detected before data is accepted.
    \item \textbf{Secure Storage on IoT Audio Device:} All user data reside on LUKS-encrypted partitions\cite{LUKS} (e.g., \texttt{/audio}, \texttt{/models}). The TPM is to anchor the Secure Boot chain of trust, it does \emph{not} retain the LUKS passphrase. After every cold start the passphrase is fetched from the API server over a mutually authenticated TLS 1.3 channel once the platform state has been successfully attested.
\end{itemize}

Whenever the underlying cryptographic stack supports it, we will enable post-quantum primitives, such as the NIST-standard Kyber key-encapsulation and Dilithium digital signatures, to initiate an early transition toward quantum-resistant security.

\subsubsection{Key Storage Implementation Details}

\begin{figure}[!ht]
    \centering
    \includegraphics[width=1\textwidth]{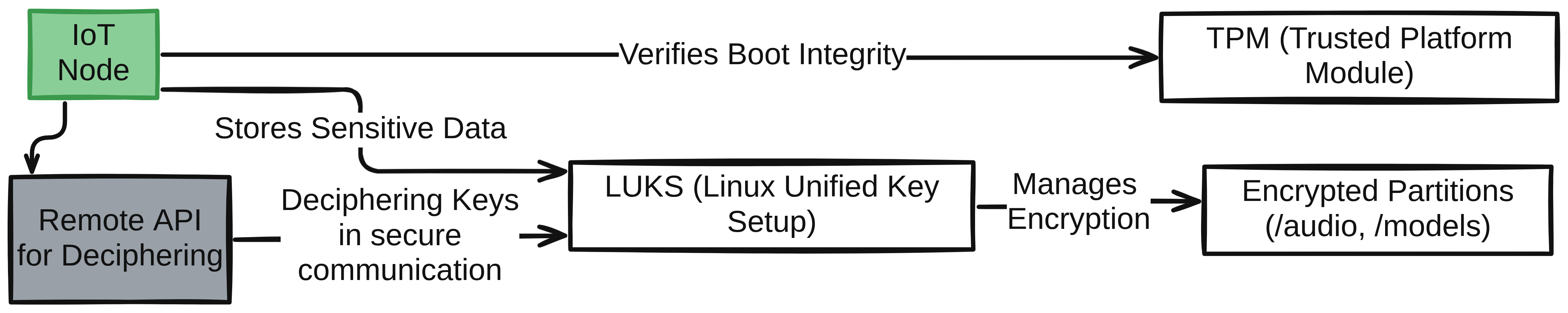}
    \caption{The diagram illustrates the API-driven LUKS unlocking workflow for IoT devices. The green box represents the IoT Node, where the unlocking process is initiated. The grey box highlights the Remote API for Deciphering.}
    \label{fig:secure_storage}
\end{figure}

The key-management scheme intentionally separates long-lived trust anchors from short-lived, high-value secrets. Device identity keys and Secure Boot measurements are sealed inside the TPM, which is resistant to physical tampering and replay. Ephemeral unlock material, on the other hand, is delivered by the cloud only after the node has proven its integrity and compliance with runtime policy. This split eliminates permanent exposure of data-at-rest keys on a stolen or rogue device while still allowing the platform to operate autonomously when network connectivity is available. An illustration of the behaviour of decrypt is shown in Fig.~\ref{fig:secure_storage}.

\begin{itemize}
    \item \textbf{Secure Boot Chain:} A hardware Root of Trust measures each stage (boot-ROM $\rightarrow$ bootloader $\rightarrow$ kernel) and extends the digests into TPM PCRs. If any measurement deviates, the kernel will not receive an unlock key, preventing the device from decrypting its data partitions.
    \item \textbf{API-Driven LUKS Unlocking on Cold Start:}
          \begin{enumerate}
              \item On power-on, the firmware measures the boot chain, the TPM merely records the hashes.
              \item Once userspace is up, an \texttt{attestation-client} daemon opens a mutually authenticated TLS 1.3 session to the API server and sends a TPM attestation quote.
              \item The API server verifies the quote, evaluates policy (device status, geofence, time window) and, if authorized, returns a \emph{one-time LUKS unlock key} inside the secure session, for added protection, this key can also be encrypted with post-quantum technology\cite{nist_pqc}.
              \item The kernel loads the key into volatile memory, decrypts the required partitions, and then zeroizes the buffer immediately.
              \item If attestation or policy evaluation fails, the server withholds the key and the encrypted partitions remain inaccessible, protecting data even if the device is lost or stolen.
          \end{enumerate}

\end{itemize}

\subsection{ML Model Security and Updates}

\begin{itemize}
    \item \textbf{Signed ML Model Updates:} ML model updates are digitally signed by a trusted authority (e.g., the API server). The IoT audio device verifies these signatures before applying any updates, preventing the injection of malicious models.
    \item \textbf{Version Control and Rollback:} The API server maintains strict version control of ML models. In case of a compromised or faulty model, a secure rollback mechanism allows deploying a previous, trusted version to the IoT audio devices.
    \item \textbf{In-transit security:} The model is protected in transit via TLS 1.3. However, to provide greater protection against emerging threats, post-quantum cryptography is employed alongside it to create a hybrid protection model.
\end{itemize}

\subsection{Certificate Management \& Backend Hardening}

End‑to‑end trust rests on two pillars:
(1) a minimal yet complete X.509 life‑cycle fully anchored in the TPM of every
IoT node, and  
(2) a defence‑in‑depth posture on the API server and its PostgreSQL backend.

\begin{itemize}
    \item \textbf{X.509 lifecycle (IoT node).}

        \begin{description}

          \item[Provisioning]
            The TPM generates (or receives at the factory) a \emph{device‑identity}
            private key and the corresponding X.509 chain (leaf, intermediate, root).
            Both key and chain are sealed in TPM NVRAM.  
            In the field, JIT‑Provisioning replaces the factory step:  
            bootstrap‑secret $\rightarrow$ CSR $\rightarrow$ CA $\rightarrow$
            certificate chain, then seal in TPM.
        
          \item[Renewal]
            Exactly 30 days before expiry, the node unseals its key, signs a fresh CSR,
            and sends it over the established mTLS session.  
            The CA returns a new certificate chain and, optionally, a freshly generated
            key. The device reseals the new material and deletes the old one.
        
          \item[Revocation]
            Backend services verify every presented certificate through OCSP stapling
            (preferred, \emph{soft‑fail with cache}), delta‑CRLs act as an offline
            fallback.  
            An out‑of‑band “emergency‑ban” channel lets operators flag a compromised
            device, the backend then blocks the certificate and forces the node back
            into the provisioning phase before it can resume service.

        \end{description}

    \item \textbf{API server \& PostgreSQL security.}

        \begin{itemize}
            \item OAuth 2.1 authentication and RBAC authorization, all requests over TLS 1.3 .
            \item Web Application Firewall in front of HTTPS endpoints, rigorous input validation blocks SQL injection, XSS, and command-injection payloads.
            \item Least-privilege roles for micro-services and database users, sensitive columns encrypted at rest (AES-256-GCM).
            \item Quarterly penetration tests and continuous vulnerability scans feed the DevSecOps backlog.
            \item Tamper-evident, signed audit logs streamed to an immutable external store to guarantee non-repudiation.
        \end{itemize}

\end{itemize}

The combined scheme preserves device authenticity, enables seamless certificate rotation, and shields the cloud backend from common web and database threats while providing verifiable traceability.

This protocol design creates a multi-layered security architecture that addresses the unique challenges of IoT audio classification devices, from the edge to the cloud.

\subsection{Physical security and tamper response}

The prototype incorporates three hardware safeguards that complement the logical controls described above.
Together, these safeguards prevent physical tampering.
They also provide forensic telemetry and, if necessary, shut down the ML subsystem and seal the data
before an attacker can extract any sensitive information.

\begin{itemize}
    \item \textbf{GPS geofencing:} A low-power GNSS receiver reports
          location on every boot and at ten-minute intervals.  If the node
          leaves its authorised polygon, the API server withholds the
          one-time LUKS\cite{LUKS} key until an operator approves the new position.

    \item \textbf{Motion sensing:} A three-axis accelerometer detects
          sustained movement or shock.  Exceeding a configurable
          threshold triggers an authenticated alert and arms the
          case-open circuit.
          
    \item \textbf{Case-open detection:} A photodiode–laser pair, backed by
          an ambient-light sensor, sits inside the enclosure.  Breaking the
          beam or a sudden rise in lux indicates that the lid has been
          removed.  The TPM’s tamper-pin is asserted, forcing an immediate
          zeroisation of volatile keys and a graceful shutdown of the ML
          inference pipeline.
          
    \item \textbf{Automated response logic:} Any tamper event drops network
          interfaces to read-only mode, snapshots the last known GPS fix,
          and stores a signed audit record in the write-once log buffer
          before power-off. Normal operation resumes only after a fresh
          remote attestation cycle.

\end{itemize}

\begin{figure}[!ht]
    \centering
    \includegraphics[width=0.8\linewidth]{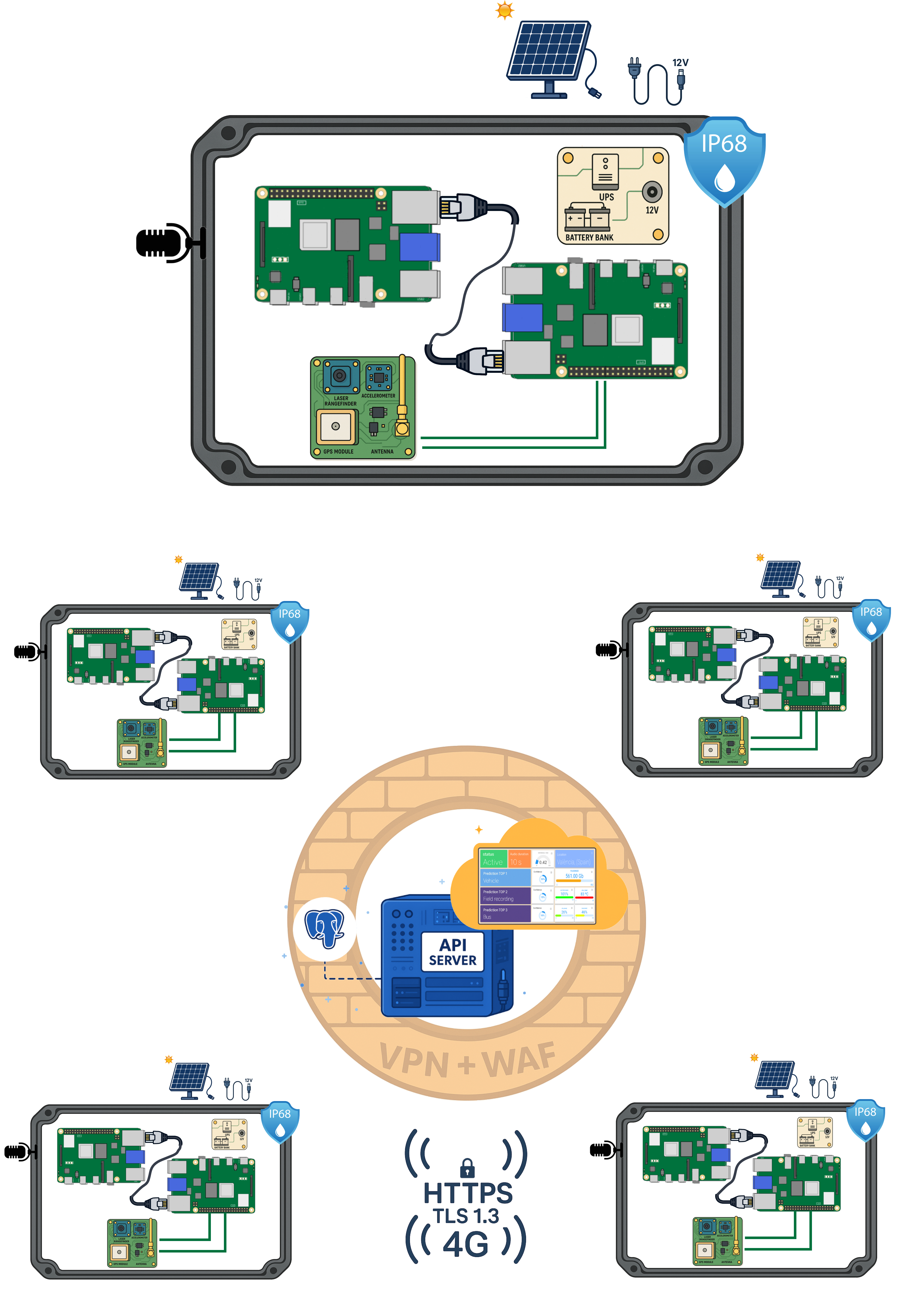}
    \caption{This diagram show the IoT architecture described for remote monitoring. Each sensing unit is housed in an IP68-certified cabinet, is powered by solar energy and a battery, and integrates electronic modules, sensors, and 4G cellular connectivity. The data collected by each station is securely transmitted via HTTPS protocols (TLS 1.3) to a central API server. This server is protected by a VPN and a web application firewall (WAF), and is used to manage, visualise and analyse the information received from multiple nodes deployed in the field.}
    \label{fig:arch}
\end{figure}

A complete wiring diagram of the prototype is provided in Fig.~\ref{fig:arch}.

\subsection{Database Backup Strategy}
\label{sec:db-backups}

To align with the 3-2-1 principle, three data copies, on two media
types, with one off-site replica~\cite{veeam321,backblaze321}, the
platform keeps encrypted backups in three distinct locations:

\begin{enumerate}
  \item \textbf{Primary SSD (hot backup).}
        A local NVMe drive stores the production PostgreSQL
        dump in a LUKS\,2 container (AES-256-XTS)~\cite{luks_best}.
        Unlock material is fetched via the attestation flow
        described in Section~\ref{sec:protocol_design}, integrity is
        verified daily with SHA-256 manifests.

  \item \textbf{Secondary HDD (cold archive).}
        Once per week the audio payload is exported, then wrapped
        with a \emph{hybrid post-quantum scheme}:
        CRYSTALS-Kyber encapsulates a session key that in turn encrypts
        the TAR archive with AES-256-GCM~\cite{kyber_hybrid,kyber_java}.
        Signatures use CRYSTALS-Dilithium~\cite{nist_pqc}.
        The disk remains offline except during the backup window,
        mitigating ransomware and supply-chain attacks~\cite{netapp_pqc}.

  \item \textbf{Cloud replica (off-site).}
        Nightly logical backups are streamed to an object-storage
        bucket protected by provider-managed AES-256-GCM at rest
        and TLS~1.3 in transit~\cite{gcp_at_rest,gcp_storage,internxt_pqc}.
        The object is further encrypted client-side with the same
        Kyber-AES hybrid key to ensure independence from the CSP.
\end{enumerate}

Each backup completes a \textbf{restore-test} every 30 days on an
isolated staging cluster, any checksum mismatch triggers an incident
per the DevSecOps playbook.

\subsection{Additional Security Recommendations}

In addition to the previously described layers of security, the following basic yet critical security recommendations should be implemented to further enhance the overall security posture of the IoT audio classification system:

\begin{itemize}
    \item \textbf{Minimal Open Ports:}
    IoT nodes and backend services should operate with all unnecessary network ports closed. Services should explicitly whitelist allowed ports and protocols, significantly reducing potential attack vectors.
    
    \item \textbf{Minimal Privilege Principle:}  
    All processes, services, and users should operate with the minimal required privileges. This applies particularly to the IoT nodes, micro-services, and database roles, ensuring compromised components have limited potential damage.
    
    \item \textbf{Secure Default Configuration:}  
    Systems and services must have secure default configurations, disabling unnecessary functionalities such as remote administrative interfaces, default accounts, or services that are not explicitly required.
    
    \item \textbf{Regular Security Updates:}  
    Both the IoT node firmware and backend server software must have a timely and automated update mechanism. Security patches for identified vulnerabilities must be applied promptly to maintain resilience against emerging threats.
    
    \item \textbf{Strong Authentication and Access Control:}  
    Strengthened authentication mechanisms, such as multi-factor authentication (MFA), should be enforced for administrative and privileged access to backend servers and sensitive operations. Role-Based Access Control (RBAC) must strictly limit permissions.
    
    \item \textbf{Logging and Monitoring:}  
    All nodes and backend infrastructure should implement comprehensive logging with continuous monitoring. Critical logs should be securely streamed off-device and protected against tampering, enhancing visibility for potential security incidents.
    
    \item \textbf{Periodic Security Audits and Assessments:}  
    Regular penetration testing, vulnerability scanning, and independent security audits should be integrated into the DevSecOps cycle, ensuring compliance with security policies and standards.
    
    \item \textbf{Device, Software and Firmware Integrity Checks:}  
    Implement regular automated checks for device and firmware integrity. Periodic remote attestation, alongside integrity checks on critical system files, ensures early detection of compromise or unauthorized modifications. To avoid impersonation, signature verification should be included in the most important parts of the developed software.
    
    \item \textbf{Secure Network Segmentation:}  
    Backend infrastructure should implement network segmentation to isolate the IoT nodes, databases, and critical services. This reduces lateral movement opportunities in the event of an intrusion. Additionally, the use of Virtual Private Networks (VPNs) is recommended to securely segment nodes and client communication, creating isolated network paths and further enhancing the confidentiality and integrity of data in transit.
    
    \item \textbf{Incident Response Plan:}  
    Clearly documented incident response procedures must be maintained. These include defined roles, responsibilities, and clear escalation paths to respond effectively and rapidly to detected incidents or breaches.

\end{itemize}

Adhering to these recommendations ensures the comprehensive security of the IoT audio classification system, further protecting sensitive audio data, the integrity of ML models, and critical infrastructure components from threats beyond those previously identified.
%!TEX root = ../main.tex
\section{Evaluation}

This section details how we will validate the security guarantees and demonstrate
that the proposed protocol solves the threats identified in Section \ref{sec:threatmodel}.
Experiments should be conducted under the production configuration described therein. This involves HTTPS over TLS 1.3 with mutual authentication, secure boot anchored to the TPM and remote delivery of an LUKS unlock key. This key is stored in volatile memory for only as long as is necessary to decrypt the volumes, after which it is immediately reset to zero.

\subsection{Experimental setup}

\begin{itemize}
    \item \textbf{IoT device:} Two Raspberry Pi 4 Model B boards
          (8 GB RAM, 512 SSD), each equipped with a TPM 2.0 module,
          GNSS receiver, three-axis accelerometer, photodiode–laser
          case-open sensor, and an independent isolated power supply.
    \item \textbf{Network:} 4G emulator with configurable RTT and packet
          loss, plus real-world field tests for validation.
    \item \textbf{Server:} 2-vCPU cloud VM,
          PostgreSQL with AES-256-GCM column encryption.
\end{itemize}

\subsection{Security effectiveness evaluation}

To demonstrate that the implemented controls withstand realistic threats,
we will carry out six complementary exercises:

\begin{itemize}

    \item \textbf{Penetration testing:} Auditors attempt spoofing,
          tampering, denial-of-service and privilege-escalation
          attacks against the IoT audio device, the 4G link, the API server,
          and the PostgreSQL backend.
          
    \item \textbf{Vulnerability scanning:} Automated tools inspect firmware,
          server code and database configuration for known CVEs and
          misconfigurations.
          
    \item \textbf{Adversarial ML attacks:} Crafted audio inputs and
          model-poisoning payloads test the effectiveness of signed-model
          enforcement and data-integrity checks.
          
    \item \textbf{Certificate and key-lifecycle audit:} From provisioning to
          revocation, we verify compliance with the X.509 workflow and the
          correct use of soft-fail OCSP caches.
          
    \item \textbf{Data-privacy assessment:} We confirm that sensitive audio
          features remain end-to-end encrypted and are never exposed in
          clear text on the device, in transit or at
          rest.

    \item \textbf{Physical assessment:} Auditors attempt to manipulate the physical
    node and launch denial-of-service attacks against it, in order to check that
    the sensors are working properly.
          
\end{itemize}

%%%%%%%%%%%%%%%%%%%%%%%%%%%%%%%%%%%%%%%%%%%%%%%%%%%%%%%%%%%%%%%%%%%%%%%%%%%%%%%%%%%%%%%%%%%%%%%%

\subsection{Formal security audit}

Static analysis will be combined with black box testing. All findings will receive CVSS scores and remediation
recommendations and residual risk statements will be provided.

\subsubsection{Static analysis}

Source code and configuration artefacts are examined without execution to
detect:
\begin{itemize}
    \item buffer overflows/underflows,
    \item injection flaws (SQL, command, deserialization),
    \item weak or deprecated cryptographic primitives,
    \item hard-coded credentials, and
    \item race conditions.
\end{itemize}

\subsubsection{Penetration testing}

Building on the previous exercise, testers simulate:
\begin{itemize}
    \item \textbf{Network attacks:} targeting client–server traffic over 4G;
    \item \textbf{Application attacks:} fuzzing API endpoints and
          model-update channels; and
    \item \textbf{Physical attacks:} attempting firmware extraction,
          side-channel analysis on the TPM, and forced reboots to bypass
          remote attestation. Success is defined as access to decrypted user
          data—mere reading of the encrypted partition does \emph{not}
          constitute a break.
\end{itemize}

\subsubsection{Vulnerability Disclosure}

All vulnerabilities are documented with impact, CVSS score and fix, the full report is released under responsible-disclosure guidelines.

\section{Conclusion}

This paper tackled the security challenges inherent in IoT audio-classification devices, now common in smart environments. Using a STRIDE-based threat model, we cataloged vulnerabilities that span the IoT node, the 4G transport, and the cloud backend. The analysis confirmed a broad attack surface and the need for a layered, edge-to-cloud defense.

We have proposed an end-to-end security protocol that provides a comprehensive theoretical security architecture. At the heart of this architecture is a mutually authenticated HTTPS channel. over TLS 1.3 that is hardened for use with intermittent cellular links. and limited hardware.  Data remains encrypted from the microcontroller unit (MCU) to the cloud, and cryptographic hashes protect its integrity. Updates to the ML model prevent the injection of adversarial code. The keying material is split between TPM-sealed anchors and single-unlock LUKS keys These are only delivered after remote attestation. On the back end, OAuth 2.1 and strict RBAC with AES-256-GCM encryption and tamper-proof audit logging protect the API server. The PostgreSQL database is protected in a similar way.

A physical-security layer, GPS geofencing, sensing, and an optical case-open detector, adds protection against theft and hardware tampering, Fig.~\ref{fig:arch} provides the full wiring diagram of the Raspberry Pi 4 prototype.  Together, these measures create a defence-in-depth posture that covers software, hardware, network, and physical domains.

\subsection*{Future work}

The evaluation of the proposed scheme will take place in the future. During implementation, security effectiveness testing should be carried out, including penetration testing, vulnerability scanning, adversarial audio attacks, and DoS stress testing. Benchmarking of CPU, memory, power, latency, and throughput performance should also be carried out. Preliminary theoretical targets suggest that robust security could be achieved without compromising real-time performance on resource-constrained audio sensors: for example, CPU overhead of less than 15\%, throughput loss of less than 10\%, and DoS resilience of at least 90\% for up to 5,000 requests per second.

Other lines of research include exploring advanced privacy-preserving techniques, such as federated learning for on-device training and differential privacy accounting. The theoretical performance impact of homomorphic encryption for secure inference will also be evaluated, alongside ethical frameworks that address user informed consent and data governance. These frameworks aim to balance utility with privacy in pervasive audio sensing applications.
%!TEX root = ../main.tex

\section*{Acknowledgements}

The participation of all the researchers in this work was funded by the Valencian Institute for Business Competitiveness (IVACE) and the FEDER funds by means of project SAVANt (IMDEEA/2025/88). The research carried out for this publication has been partially funded by the project STARRING-NEURO (PID2022-137048OA-C44) funded by the Agencia Estatal de Investigación (Spanish State Research Agency) and the Europan Union MCIN/AEI/10.13039/501100011033/ FEDER, UE.

\newpage

\printindex % Print the index

\bibliographystyle{plain}
\bibliography{refs}

@article{Vardakis2024Review,
  title={Review of Smart-Home Security Using the Internet of Things},
  author={Vardakis, George and Hatzivasilis, George and Koutsaki, Eleftheria and Papadakis, Nikos},
  journal={Electronics},
  volume={13},
  number={16},
  pages={3343},
  year={2024},
  publisher={MDPI},
  doi={10.3390/electronics13163343}
}

@article{Wang2023Security,
  title={Security Issues on Industrial Internet of Things: Overview and Challenges},
  author={Wang, Maoli and Sun, Yu and Sun, Hongtao and Zhang, Bowen},
  journal={Computers},
  volume={12},
  number={12},
  pages={256},
  year={2023},
  publisher={MDPI},
  doi={10.3390/computers12120256}
}

@article{LoScudo2023Audio,
  title={Audio-based anomaly detection on edge devices via self-supervision and spectral analysis},
  author={Lo Scudo, Fabrizio and Ritacco, Ettore and Caroprese, Luciano and Manco, Giuseppe},
  journal={Journal of Intelligent Information Systems},
  volume={61},
  number={3},
  pages={765--793},
  year={2023},
  publisher={Springer US}
}

@article{Al-Khalli2023Real-Time,
  title={Real-Time Detection of Intruders Using an Acoustic Sensor and Internet-of-Things Computing},
  author={Al-Khalli, Najeeb and Alateeq, Saud and Almansour, Mohammed and Alhassoun, Yousef and Ibrahim, Ahmed B. and Alshebeili, Saleh A.},
  journal={Sensors},
  volume={23},
  number={13},
  pages={5792},
  year={2023},
  publisher={MDPI},
  doi={10.3390/s23135792}
}

@Article{Yang2025RealTimeASR,
AUTHOR = {Yang, Hongyu and Dong, Rou and Guo, Rong and Che, Yonglin and Xie, Xiaolong and Yang, Jianke and Zhang, Jiajin},
TITLE = {Real-Time Acoustic Scene Recognition for Elderly Daily Routines Using Edge-Based Deep Learning},
JOURNAL = {Sensors},
VOLUME = {25},
YEAR = {2025},
NUMBER = {6},
ARTICLE-NUMBER = {1746},
URL = {https://www.mdpi.com/1424-8220/25/6/1746},
PubMedID = {40292891},
ISSN = {1424-8220},
DOI = {10.3390/s25061746}
}

@article{Ammar2018InternetOT,
  title={Internet of Things: A survey on the security of IoT frameworks},
  author={Mahmoud Ammar and Giovanni Russello and Bruno Crispo},
  journal={Journal of Information Security and Applications},
  year={2018},
  volume={38},
  pages={8-27}
}

@incollection{Kumar2021Survey,
  title     = {A Survey of Lightweight Cryptography for Power-Constrained IoT Devices: Security Challenges and Issues},
  author    = {Kumar, Sunil and Kumar, Dilip},
  booktitle = {Green Engineering and Technology},
  editor    = {Jena, Om Prakash and Tripathy, Ashok R. and Polkowski, Zdzislaw},
  publisher = {CRC Press},
  year      = {2021},
  pages     = {293--313},
  isbn      = {9781003176275}
}

@article{Turchet2020Internet,
  title={The Internet of Audio Things: State of the Art, Vision, and Challenges},
  author={L. Turchet and G. Fazekas and M. Lagrange and H. S. Ghadikolaei and C. Fischione},
  journal={IEEE Internet of Things Journal},
  volume={7},
  number={10},
  pages={10233-10249},
  year={2020},
  publisher={IEEE},
  doi={10.1109/JIOT.2020.2997047}
}

@misc{Chen2024Adversarial,
      title={Holistic Adversarial Robustness of Deep Learning Models}, 
      author={Pin-Yu Chen and Sijia Liu},
      year={2023},
      eprint={2202.07201},
      archivePrefix={arXiv},
      primaryClass={cs.LG},
      url={https://arxiv.org/abs/2202.07201}, 
}

@article{Brown2024FirmwareSecurity,
   author  = {W. M. A. B. Wijesundara and Joong-Sun Lee and Dara Tith and Eleni Aloupogianni and Hiroyuki Suzuki and Takashi Obi},
  title   = {Security{\textendash}Enhanced Firmware Management Scheme for Smart Home
             {IoT} Devices Using Distributed Ledger Technologies},
  journal = {International Journal of Information Security},
  volume  = {23},
  pages   = {1927--1937},
  year    = {2024},
  doi     = {10.1007/s10207-024-00827-x}
}

@inproceedings{Jones2023ThreatModeling,
  author    = {Christian Ellerhold and Johann Schnagl and Thomas Schreck},
  title     = {Enterprise Cyber Threat Modeling and Simulation of Loss Events for
               Cyber Risk Quantification},
  booktitle = {Proceedings of the 2023 Cloud Computing Security Workshop ({CCSW}\'23)},
  pages     = {17--29},
  year      = {2023},
  doi       = {10.1145/3605763.3625244}
}

@INPROCEEDINGS{Zhao2024PhysicalSecurity,
  author={Zhao, Hanning and Silverajan, Bilhanan},
  booktitle={2022 Thirteenth International Conference on Ubiquitous and Future Networks (ICUFN)}, 
  title={User-Centered Design to Enhance IoT Cybersecurity Awareness of Non-Experts in Smart Buildings}, 
  year={2022},
  volume={},
  number={},
  pages={369-371},
  doi={10.1109/ICUFN55119.2022.9829563}}

@INPROCEEDINGS{Kim2025LightweightDTLS,
  author={Haroon, Asma and Akram, Sana and Shah, Munam Ali and Wahid, Abdul},
  booktitle={2017 IEEE 86th Vehicular Technology Conference (VTC-Fall)}, 
  title={E-Lithe: A Lightweight Secure DTLS for IoT}, 
  year={2017},
  volume={},
  number={},
  pages={1-5},
  doi={10.1109/VTCFall.2017.8288362}}

@techreport{Liu2025PostQuantum,
  author      = {Alagic, G. and et al.},
  title       = {Status Report on the Third Round of the NIST Post-Quantum Cryptography Standardization Process},
  institution = {National Institute of Standards and Technology},
  number      = {NIST IR 8413},
  year        = {2022},
  doi         = {10.6028/NIST.IR.8413}
}

@article{Taylor2021Homomorphic,
author = {Ganesh Kumar Mahato and Swarnendu Kumar Chakraborty},
title = {A Comparative Review on Homomorphic Encryption for Cloud Security},
journal = {IETE Journal of Research},
volume = {69},
number = {8},
pages = {5124--5133},
year = {2023},
publisher = {Taylor \& Francis},
doi = {10.1080/03772063.2021.1965918},

URL = {   
        https://doi.org/10.1080/03772063.2021.1965918

},
eprint = {    
        https://doi.org/10.1080/03772063.2021.1965918

}
}

@InProceedings{Garcia2025Privacy,
author="Bourr{\'e}e, Jade Garcia
and Lautraite, Hadrien
and Gambs, S{\'e}bastien
and Tredan, Gilles
and Le Merrer, Erwan
and Rottembourg, Beno{\^i}t",
editor="Ribeiro, Rita P.
and Pfahringer, Bernhard
and Japkowicz, Nathalie
and Larra{\~{n}}aga, Pedro
and Jorge, Al{\'i}pio M.
and Soares, Carlos
and Abreu, Pedro H.
and Gama, Jo{\~a}o",
title="P2NIA: Privacy-Preserving Non-iterative Auditing",
booktitle="Machine Learning and Knowledge Discovery in Databases. Research Track",
year="2026",
publisher="Springer Nature Switzerland",
address="Cham",
pages="259--275",
isbn="978-3-032-06096-9"
}

@misc{Miller2025AIAnomaly,
      title={Detection of anomalies amongst LIGO's glitch populations with autoencoders}, 
      author={Paloma Laguarta and Robin van der Laag and Melissa Lopez and Tom Dooney and Andrew L. Miller and Stefano Schmidt and Marco Cavaglia and Sarah Caudill and Kurt Driessens and Jöel Karel and Roy Lenders and Chris Van Den Broeck},
      year={2023},
      eprint={2310.03453},
      archivePrefix={arXiv},
      primaryClass={astro-ph.IM},
      url={https://arxiv.org/abs/2310.03453}, 
}

@article{Smith2024EdgeAI,
  author  = {Heath Smith and James Seekings and Mohammadreza Mohammadi and
             Ramtin Zand},
  title   = {Realtime Facial Expression Recognition: Neuromorphic Hardware vs.
             Edge {AI} Accelerators},
  journal = {arXiv preprint},
  eprint  = {2403.08792},
  year    = {2024}
}

@misc{luks_best,
  author    = {Gilles 'SO- stop being evil'},
  title     = {Best practice to backup a LUKS encrypted device},
  year      = {2014},
  howpublished = {\url{https://unix.stackexchange.com/questions/101888/best-practice-to-backup-a-luks-encrypted-device}},
  urldate   = {2025-07-14}
}

@misc{kyber_hybrid,
  author    = {Bill Buchanan},
  title     = {Hybrid Encryption with Kyber (ML-KEM) and AES},
  howpublished = {\url{https://asecuritysite.com/kyber/go_hybrid3}},
  note      = {Accessed 2025-07-14}
}

@misc{kyber_java,
  author    = {Pathum Hewage},
  title     = {Using CRYSTALS-Kyber KEM for Hybrid Encryption with Java},
  year      = {2024},
  howpublished = {\url{https://medium.com/@hwupathum/using-crystals-kyber-kem-for-hybrid-encryption-with-java-0ab6c70d41fc}},
  urldate   = {2025-07-14}
}

@misc{nist_pqc,
  institution = {National Institute of Standards and Technology},
  title     = {NIST Releases First 3 Finalized Post-Quantum Encryption Standards},
  year      = {2024},
  howpublished = {\url{https://www.nist.gov/news-events/news/2024/08/nist-releases-first-3-finalized-post-quantum-encryption-standards}},
  urldate   = {2025-07-14}
}

@misc{netapp_pqc,
  institution = {NetApp},
  title     = {Post-Quantum Cryptography | NetApp},
  year      = {2025},
  howpublished = {\url{https://www.netapp.com/cyber-resilience/post-quantum-cryptography/}},
  urldate   = {2025-07-14}
}

@misc{veeam321,
  author    = {Anthony Spiteri},
  title     = {3-2-1-1-0 Golden Backup Rule},
  year      = {2021},
  howpublished = {\url{https://community.veeam.com/blogs-and-podcasts-57/3-2-1-1-0-golden-backup-rule-569}},
  urldate   = {2025-07-14}
}

@misc{backblaze321,
  author    = {Andy Klein},
  title     = {Why the 3-2-1 Backup Strategy is the Best},
  year      = {2024},
  howpublished = {\url{https://www.backblaze.com/blog/the-3-2-1-backup-strategy/}},
  urldate   = {2025-07-14}
}

@misc{gcp_at_rest,
  institution = {Google Cloud},
  title     = {Default Encryption at Rest},
  year      = {2024},
  howpublished = {\url{https://cloud.google.com/docs/security/encryption/default-encryption}},
  urldate   = {2025-07-14}
}

@misc{gcp_storage,
  institution = {Google Cloud},
  title     = {Standard Cloud Storage Encryption},
  year      = {2025},
  howpublished = {\url{https://cloud.google.com/storage/docs/encryption/default-keys}},
  urldate   = {2025-07-14}
}

@misc{internxt_pqc,
  author    = {Internxt Blog Team},
  title     = {The 11 Best Encrypted Cloud Storage Solutions 2025},
  year      = {2024},
  howpublished = {\url{https://blog.internxt.com/encrypted-cloud-storage/}},
  urldate   = {2025-07-14}
}

@misc{NIST,
  author = {Chandramouli, Ramaswamy and Butcher, Zack},
  title  = {Guidelines for API Protection for Cloud-Native Systems},
  howpublished = {\url{https://doi.org/10.6028/NIST.SP.800-228}},
  year   = {2025},
  note   = {[Accessed 2025-07-17]}
}

@article{STRIDE,
title = {Specification, detection, and treatment of STRIDE threats for software components: Modeling, formal methods, and tool support},
journal = {Journal of Systems Architecture},
volume = {117},
pages = {102073},
year = {2021},
issn = {1383-7621},
doi = {https://doi.org/10.1016/j.sysarc.2021.102073},
url = {https://www.sciencedirect.com/science/article/pii/S1383762121000631},
author = {Quentin Rouland and Brahim Hamid and Jason Jaskolka},

}

@Article{LUKS,
AUTHOR = {Cano-Quiveu, German and Ruiz-de-clavijo-Vazquez, Paulino and Bellido, Manuel J. and Juan-Chico, Jorge and Viejo-Cortes, Julian and Guerrero-Martos, David and Ostua-Aranguena, Enrique},
TITLE = {Embedded LUKS (E-LUKS): A Hardware Solution to IoT Security},
JOURNAL = {Electronics},
VOLUME = {10},
YEAR = {2021},
NUMBER = {23},
ARTICLE-NUMBER = {3036},
URL = {https://www.mdpi.com/2079-9292/10/23/3036},
ISSN = {2079-9292},

DOI = {10.3390/electronics10233036}
}

\end{document}